\documentclass[10pt]{article}

\usepackage[letterpaper,margin=1in]{geometry}
\usepackage[T1]{fontenc}
\usepackage{times}
\usepackage{amsmath}
\usepackage{amssymb}
\usepackage{booktabs}
\usepackage{graphicx}
\usepackage{siunitx}
\usepackage[hidelinks]{hyperref}
\usepackage[numbers,sort&compress]{natbib}
\usepackage{caption}
\graphicspath{{figures/}}

\title{\vspace{-2.5em}Misspecification in amortized X-ray spectral inference:\\
a detection benchmark, a gain shift three schemes cannot see,\\
and marginalizing it out}

\author{
  Karan Akbari\,\thanks{\href{https://orcid.org/0009-0005-0550-4018}{ORCID 0009-0005-0550-4018}; \texttt{karanakbari14@gmail.com}}\\
  St.\ Xavier's College, Mumbai, India
}
\date{}

\begin{document}
\maketitle

\begin{abstract}
\noindent
Amortized neural posterior estimation fits X-ray spectra in milliseconds, and pipeline use will lean on post-hoc checks to catch detector systematics the simulator misses. We measure what those checks see, on simulated spectra from real XMM-Newton EPIC-pn and NICER responses. Of four misspecification families, an unmodelled 6.4\,keV line and partial covering are catchable per spectrum (AUC 0.97 and 0.84). A wrong continuum family reaches 0.66. One flow passed every recovery check yet under-covered (coverage deviation 0.114), and a converged NICER flow (0.011) fails simulation-based calibration on all five parameters. Recovery metrics do not certify calibration. Three schemes catch the line: the posterior-predictive check, importance sampling against the exact Poisson likelihood (effective sample size, 0.68--0.82), and the nested-sampling evidence (count-controlled, $-67$ to $-892$ nats). None catches the 3 per cent gain shift, 2.4 to 14.4 times the EPIC-pn energy-scale accuracy (AUC 0.44--0.54; 36-cell means 0.50 and 0.49 on both responses; paired evidence $+0.33 \pm 1.37$ nats). The effective sample size stays null-consistent from 0.1 to 10 per cent, realistic amplitudes included. The shift still biases the photon index by $+0.02$ (0.03 to 0.11$\sigma$) at both count levels. Retraining with a per-simulation nuisance gain adds no inference-time cost, and the gain posterior returns the prior. The bias survives, and exact gain-marginalized nested sampling shows the flow capturing at most 0.163 of its gain constraint (continuous-prior convention) at high counts. Marginalization over posited calibration nuisances is the working defence for X-ray SBI at catalog scale; no check we ran supplies a warning.

\end{abstract}

\section{Introduction}\label{sec:intro}

Amortized simulation-based inference (SBI) is arriving in X-ray spectroscopy. A neural posterior estimator (NPE) trains once on simulated spectra and then returns a posterior for any new spectrum in milliseconds, where a per-source sampler takes minutes to hours, and implementations aimed at pipeline use already exist \citep{barret2024_i,dupourque2025_ii}. The speed has a price. The flow only knows spectra its training simulator can produce, so a detector systematic missing from that simulator biases every posterior the flow returns, with no per-source warning and at catalog scale.

We measure this on two real instrument responses, the XMM-Newton EPIC-pn and the NICER XTI (Section~\ref{sec:detection}). An injected 6.4\,keV emission line is the easy case, and the posterior-predictive check catches it from ${\sim}1000$ counts up. A 3 per cent detector gain shift slips through. It biases the fitted photon index while every per-spectrum detection score stays at chance on both responses, and a paired nested-sampling comparison on matched spectra puts the gain's evidence signal at $+0.3 \pm 1.4$ nats, consistent with zero. We also build a count-controlled evidence residual, reading each spectrum's $\log Z$ against the clean $\log Z$--counts trend, and its medium-counts gain cell sits at $+13$ nats, an artifact of the regression itself.

So we widen the question, and put three schemes on the same misspecification families, at the same amplitudes and count levels, on one EPIC-pn response: the NPE's own posterior-predictive discrepancy, importance weights computed under the exact Poisson likelihood, and the evidence from nested sampling with its sampling error measured. Each scheme has its own draw of spectra, and the table captions carry the sample sizes. \citet{barret2026_iii} propose the second scheme, importance reweighting of the NPE posterior with the effective sample size as a quality diagnostic, and they explicitly set aside the question of whether that diagnostic notices a systematic the simulator does not contain. We run that test, and we add the Pareto-smoothed $\hat{k}$ \citep{vehtari2024psis}. Every scheme sees the line. The gain shift survives all three, and an amplitude sweep shows that is not for lack of power, since the effective-sample-size (ESS) detector stays at chance from 1 to 10 per cent injected gain. The count-controlled evidence residual goes too. We quantify that artifact and confirm it, but the construction needs the clean trend, which an analyst fitting a real spectrum does not have, so it survives only as a benchmark instrument. So the blindness sits with the systematic and with the five-parameter model that reabsorbs it, and Section~\ref{sec:mechanism} works out why.

The 3 per cent shift is a deliberate stress test. The EPIC-pn absolute energy-scale accuracy is ${\pm}12.5$\,eV in imaging modes \citep{smith2025epiccal}, which is 0.2 per cent at 6\,keV and 1.25 per cent at 1\,keV; NICER's is ${\sim}5$\,eV \citep{markwardt2023nicercal}, or 0.083 per cent at 6\,keV and 0.5 per cent at 1\,keV. A slope-only gain is pinned at the top of the band, where the same ${\pm}12.5$\,eV is 0.125 per cent at 10\,keV. The 1\,keV figure is the loosest of the three EPIC-pn values, and 0.1 to 0.3 per cent is the range we treat as realistic below. Against the 1 and 6\,keV values above, our injected shift exceeds the EPIC-pn figure by a factor of 2.4 to 14.4 and the NICER figure by 6 to 36, and against the realistic band itself it is 10 to 30 times too large. At the realistic 0.1 to 0.3 per cent level the shift is only harder to see, and we confirm the invisibility there directly.

So the systematic has to be modelled upstream, and we draw a nuisance gain in every training simulation, retrain, and marginalize at inference, which trains the flow through the whole miscalibration family and adds nothing to the per-spectrum inference cost. The posterior widths do not move, and for a ${\pm}5$ per cent gain prior they are not supposed to (Section~\ref{sec:fixbias}). The photon-index bias a true 3 per cent shift induces, about $+0.02$, is physical and survives marginalization, because at these counts the data carry almost no gain information and the gain posterior returns the prior. And on the few bright spectra where gain information does exist, an exact gain-marginalized nested-sampling run shrinks the gain posterior to 0.73 to 0.86 of the prior while the amortized flow captures at most 0.163 of that shrink, on the continuous-prior convention. Marginalization is still the right default, and Section~\ref{sec:fixns} measures where it stops working.

Section~\ref{sec:setup} describes the simulator, the responses, and the misspecification families, Section~\ref{sec:detection} measures which of those families a per-spectrum check can catch, and Section~\ref{sec:calibration} audits the flows themselves with rank and coverage tests. Section~\ref{sec:isess} runs the deferred importance-sampling test on gain shifts, and Section~\ref{sec:matrix} assembles the cross-scheme matrix with corrected nested-sampling error bars. Section~\ref{sec:fix} builds and validates the gain-marginalized flow, Section~\ref{sec:limits} pins the results at realistic amplitude, and Section~\ref{sec:mechanism} gives the geometric reading. Section~\ref{sec:conclusions} concludes.

\section{Setup}\label{sec:setup}

All simulation uses \textsc{jaxspec} \citep{dupourque2024_jaxspec} with two real instrument responses: the XMM-Newton EPIC-pn response bundled with the NGC~7793 ULX-4 observation \citep{quintin2021}, grouped to 102 channels, and the public on-axis NICER XTI response (ARF \texttt{nixtiaveonaxis20170601}, RMF \texttt{nixtiref20170601}, 969 channels, 0.3--10\,keV). The model is \texttt{tbabs}$\times$(\texttt{powerlaw}+\texttt{blackbody}) with five free parameters: $N_{\rm H}$, the photon index $\Gamma$, the power-law normalization, $kT_{\rm bb}$, and the blackbody normalization. Priors follow Table~1 of \citet{barret2024_i}, with the two log-uniform normalization windows shifted down, and counts are set through an effective exposure that rescales the transfer matrix, so the three regimes land at median total counts near 100, 1000, and 10000 over the prior. We call them faint, medium, and bright. The medians hide a wide spread: individual medium-level spectra span roughly two orders of magnitude (98 to 6492 counts in the nested-sampling benchmark's own draw), so every population statistic below is population-averaged.

Four misspecification families perturb the simulator, each on a four-point strength grid; the weakest grid point sits next to the clean model and acts as the control, though for B2 and B3 that weakest point already separates from clean (Table~\ref{tab:auc}). B1 adds a narrow unmodelled Gaussian emission line at 6.4\,keV ($\sigma = 0.05$\,keV) and spans its normalization over $5\times10^{-6}$, $2\times10^{-5}$, $8\times10^{-5}$, and $3\times10^{-4}$\,photons\,cm$^{-2}$\,s$^{-1}$; the nested-sampling rows use the strongest of the four, an intrinsic line flux added to the continuum before absorption and response folding. B2 swaps \texttt{tbabs} for a partial-covering absorber and spans covering fraction 0.9, 0.7, 0.5, and 0.3. B3 replaces the power law with a thermal-bremsstrahlung template $M(E)=K\,E^{-1}\exp(-E/kT)$ and spans $kT = 10$, 5, 3, and 1.5\,keV. B4 is the gain shift, swept at 0.5, 1, 2, and 3 per cent: the response's unfolded energy grid is rescaled in place by a gain factor $g$ (3 per cent means $g=1.03$), slope only, zero offset, so there is no constant term for a fit to pick up. At 3 per cent the shift moves a 6.4\,keV feature by about 190\,eV on EPIC-pn and about 1.3 resolution elements on NICER.

One neural spline flow is trained per count level with the \textsc{sbi} package \citep{boelts2025} on 50\,000 simulations, conditioned through a one-dimensional CNN embedding of two convolution blocks, 16 and 32 channels, kernel 5, into a 20-dimensional summary. The flow carries five spline transforms with 50 hidden features and 10 bins, and training is capped at 150 epochs with patience-20 early stopping. The flows stay level-specific because coverage is exposure-dependent, and a single flow amortized across regimes would confound a calibration study. Every likelihood-based check in this paper uses the same exact Poisson likelihood, pinned by a unit test to $10^{-9}$ agreement with the one nested sampling evaluates. Nested sampling runs with \textsc{UltraNest} \citep{buchner2021,buchner2023} on a 76-spectrum subsample (56 clean across the three levels, plus B1 and B4 spectra at medium and bright), capped at $1.2\times10^{5}$ likelihood calls per spectrum; 12 rows hit the cap, none of them line spectra, and the largest \textsc{UltraNest} internal $\log Z$ error on a capped row is 0.76 nats. That internal estimate is not the bootstrap error bar we use in Section~\ref{sec:matrix}. Two later sets of runs sit outside this benchmark and outside its cap: the 12 gain pairs of Section~\ref{sec:matrix}, which run to a larger call budget, and the uncapped gain-marginalized runs of Section~\ref{sec:fixns}. Per spectrum, nested sampling is 8800 to 13000 times slower than the flow.

The detection benchmark of Section~\ref{sec:detection} runs three detectors. D1 is a per-spectrum posterior-predictive check: parameter draws from the flow are folded through the response, Poisson-realized, and scored with $\chi^2$ and KS discrepancies on counts, descended from the cumulative-QQ model discovery of \citet{buchner2014}. D2 is a per-spectrum out-of-distribution distance in the flow's learned embedding space, a single-spectrum adaptation of the population-level summary-space check of \citet{schmitt2023}. D3 is the marginal, population-level classifier two-sample test \citep{lopezpaz2017c2st}: one classifier per benchmark cell, trained on the flow's embedding features to separate a whole clean population from a misspecified one. Sections~\ref{sec:isess}--\ref{sec:fix} run on the EPIC-pn response and on families B1 and B4. The detection benchmark covers all four families on EPIC-pn, and its NICER replication covers B1, B3 and B4 on all three detectors.

\section{Detection}\label{sec:detection}

\begin{figure}[t]
\centering
\includegraphics[width=0.96\linewidth]{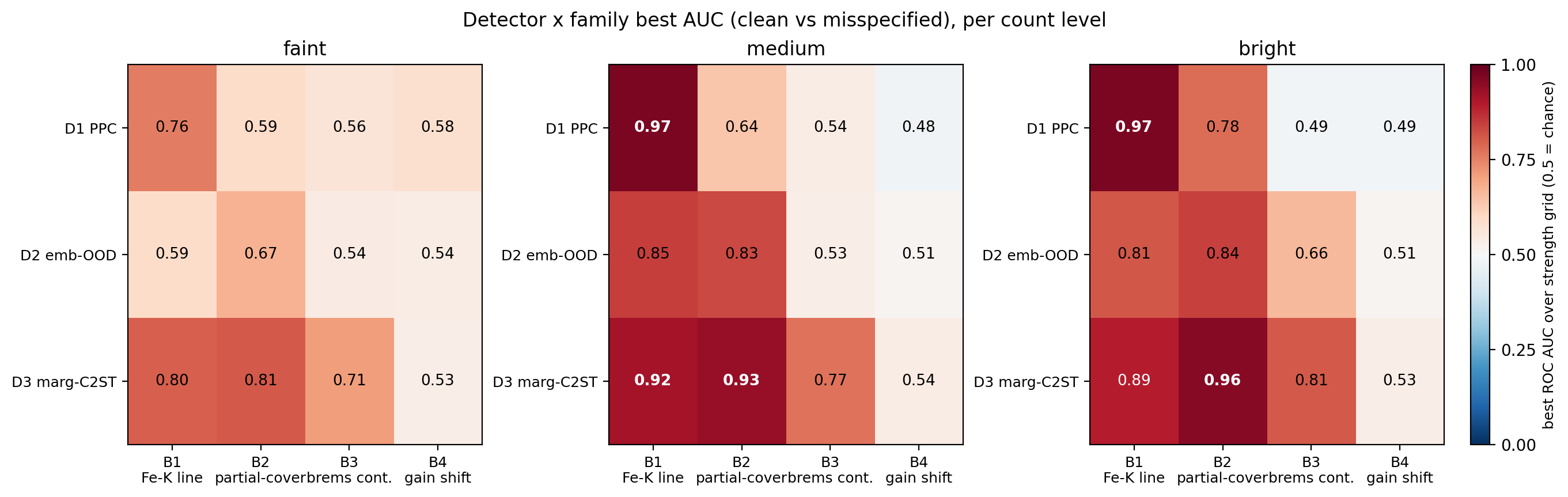}
\caption{Detection ROC AUC for the three detectors (rows) across the four
misspecification families (columns), one panel per count level, each cell taken over
the four-point strength grid. Brighter cells are more detectable, and $0.5$ is chance.
Table~\ref{tab:auc} carries the per-cell better of D1 and D2 from this grid, together
with the D3 row. The B4 gain-shift column stays at chance for all three detectors at
every count level. D3 is the supervised population statistic and carries a non-$0.5$
control-cell floor (Section~\ref{sec:detection}).}
\label{fig:auc_grid}
\end{figure}

The detection benchmark is a 144-cell ROC grid (four families $\times$ four strengths
$\times$ three count levels $\times$ three detectors), summarized in
Table~\ref{tab:auc} and plotted per detector in Fig.~\ref{fig:auc_grid}. The per-spectrum detectors
D1 and D2 match the practical case: an analyst sees one spectrum at a time,
without the labelled clean-versus-misspecified population a population statistic
needs. D3 is that population statistic. It is a supervised population-separability
statistic, not a per-spectrum trust score, so we report it in a separate column and
read it against its control-cell floor of $\approx 0.66$ cross-validated accuracy;
the per-spectrum conditional version was pathological against the over-confident
posteriors of Section~\ref{sec:calibration}, so D3 answers a population-level question. Each AUC is computed on
100 misspecified versus 200 clean spectra, a per-cell standard error near $0.04$, so a
difference between two cells carries a standard error of $\sqrt{2}\times0.04 = 0.057$
and differences below ${\sim}0.11$ are within noise; the contrasts we draw are the large
per-family differences and the 36-cell B4 aggregate, not fine per-cell orderings.

\begin{table}[t]
\centering
\caption{Best ROC AUC per (count level, family) among the per-spectrum detectors
(D1/D2), maximized over the strength grid, with the winning detector named in
parentheses wherever D1 and D2 separate by more than $0.05$; in the unlabelled cells the two
sit within $0.05$ of each other. $0.5$ is chance. The D3 columns are the marginal C2ST of
Section~\ref{sec:setup}, a supervised population-separability
statistic, and they are not per-spectrum trust scores. Their control-cell floor is about $0.66$
cross-validated accuracy, which is a different scale from the AUC entries;
in AUC it spans $0.33$--$0.54$ for B1/B4 but sits at $0.61$--$0.63$ for B2 and
$0.68$--$0.77$ for B3, whose weakest grid points are not exactly clean. The bright rows run on
the production flow Section~\ref{sec:calibration} finds miscalibrated. Fig.~\ref{fig:auc_grid}
plots the full three-detector grid these entries are read from.}
\label{tab:auc}
\small
\begin{tabular}{l cccc c cccc}
\toprule
 & \multicolumn{4}{c}{Per-spectrum (D1/D2)} & & \multicolumn{4}{c}{Population (D3)}\\
\cmidrule(lr){2-5}\cmidrule(lr){7-10}
Level & B1 & B2 & B3 & B4 & & B1 & B2 & B3 & B4\\
\midrule
faint  & 0.76 (D1) & 0.67 (D2) & 0.56 & 0.58 & & 0.80 & 0.81 & 0.71 & 0.53\\
medium & 0.97 (D1) & 0.83 (D2) & 0.54 & 0.51 & & 0.92 & 0.93 & 0.77 & 0.54\\
bright & 0.97 (D1) & 0.84 (D2) & 0.66 (D2) & 0.51 & & 0.89 & 0.96 & 0.81 & 0.53\\
\bottomrule
\end{tabular}
\end{table}

The posterior-predictive check catches the Fe-K line (B1), with D1
reaching AUC $0.97$ at and above $\sim$1000 counts, because the line adds localized
counts in the $5.8$--$7.0$\,keV window that the posterior-predictive replicates
cannot reproduce. At $\sim$100 counts shot noise buries it and the best
per-spectrum AUC falls to $0.76$. A missed line biases the
continuum. The inferred photon index is pulled softer, and the signed bias grows
with counts, from $+0.09$ on the mean at medium (median $+0.10$) to $+0.20$ at
bright (median $+0.26$) for the strongest line. At faint
the signed shift is scatter-dominated and slightly negative ($-0.07$).

Partial covering (B2) is where the embedding detector helps. D2 climbs from $0.67$
to $0.84$ with counts while D1 lags (Fig.~\ref{fig:auc_grid}), because a covering fraction reshapes the whole
soft continuum, a global distortion the embedding sees better than the channel-wise
check. The wrong
continuum family (B3) is harder. The population statistic D3 separates the
bremsstrahlung family from clean at every level (AUC $0.71$--$0.81$), and it already reaches
$0.68$--$0.77$ at the weakest grid point, so the separability comes from the family
itself: the population classifier sees the wrong continuum
family at every $kT$, including the most power-law-like one. The per-spectrum
detectors stay near chance almost everywhere, because at the $kT$ where bremsstrahlung looks most
power-law-like the 5-parameter model absorbs the difference into $N_{\rm H}$,
$\Gamma$, and the blackbody. The exception is bright, where D2 reaches $0.66$ at
its best grid point, four per-cell standard errors above chance and the only
per-spectrum detection this family gets.

The gain shift (B4) is the main negative result. None of the three detectors
flags it at any count level or gain strength. Across the $0.5$--$3\%$ grid all 36
B4 cells, pooling both per-spectrum detectors with the population statistic, span
AUC $0.43$--$0.58$ with mean $0.50$, flat in counts and flat in gain.
The grouped EPIC-pn channels near Fe-K are $175$\,eV wide, so the $3\%$ shift moves
the line by about one channel and channel coarseness alone could hide much of the
grid there. NICER samples the band with
$10$\,eV channels, so coarse sampling is not the issue on that side, but its measured
resolution at the line, the FWHM of its own 6.4\,keV redistribution, is about
$145$\,eV, which puts the $3\%$ shift at about $1.3$ resolution elements there too,
and its 36 B4 cells still sit at chance (mean AUC $0.49$): a rescaled continuum has
no localized feature for the resolution to act on, unlike the line. For a single
unlabelled spectrum, the posterior-predictive check flags
lines and the embedding distance flags partial covering; the gain shift gets past both
(Fig.~\ref{fig:money}\textit{b}).

\begin{figure}[t]
\centering
\includegraphics[width=0.98\linewidth]{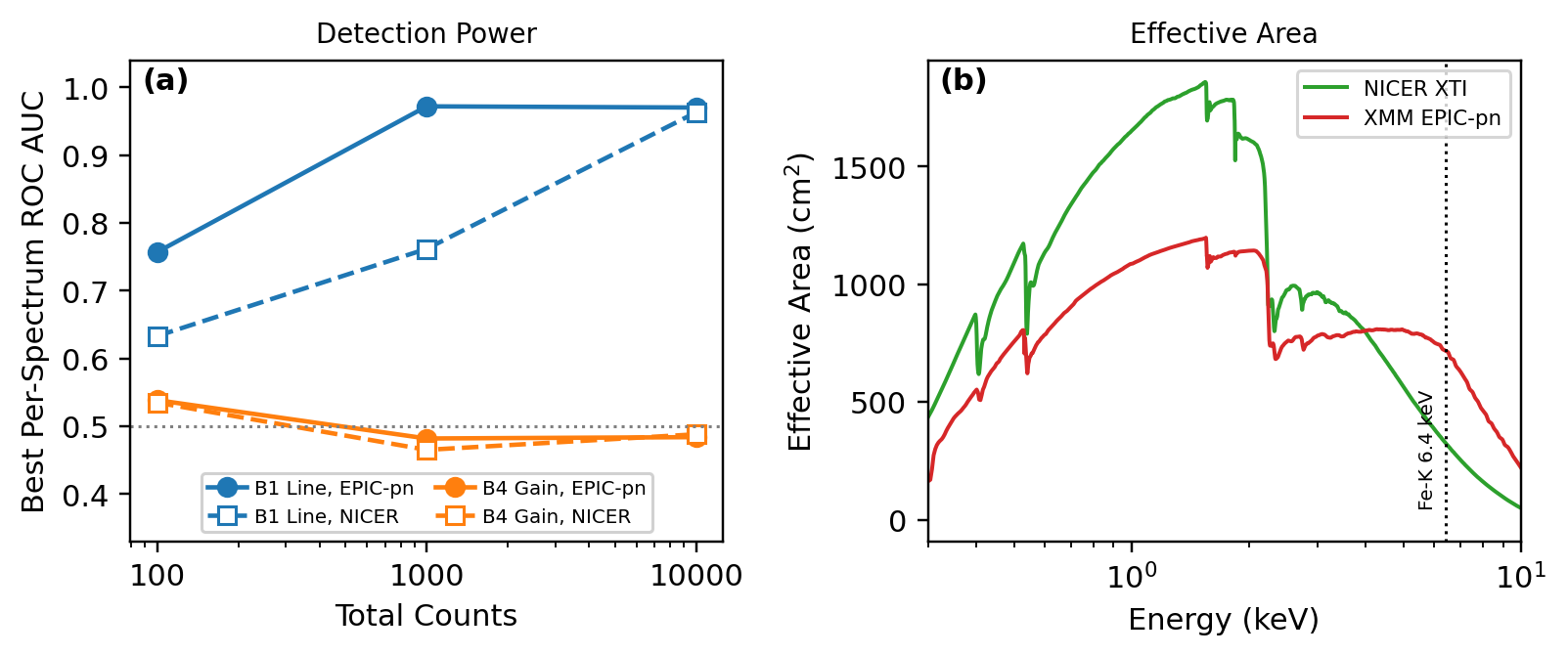}
\caption{Replication on NICER. \textit{(a)} Best per-spectrum ROC AUC against total counts
for the Fe-K line (B1, blue) and the 3 per cent gain shift (B4, orange), on EPIC-pn
(filled, solid) and NICER (open, dashed), with the dotted line at chance. The gain shift
stays at chance on both responses. The line is caught on both and reaches AUC
$0.96$--$0.97$ at $\sim$10000 counts, and it is weaker on NICER at lower counts.
\textit{(b)} Effective area against energy for the two responses, with the 6.4\,keV Fe-K
line marked. The curves cross near 4\,keV, so at the line NICER sits below EPIC-pn and
about six times below its own 1.5\,keV peak, and at matched total counts fewer photons
land near the line (Section~\ref{sec:detection}).}
\label{fig:nicer}
\end{figure}

Running the same benchmark on the NICER XTI response reproduces both results
(Fig.~\ref{fig:nicer}), and the
gain shift stays at chance there too: its 36 B4 cells, the same three-detector pool, span
AUC $0.41$--$0.57$ with mean $0.49$. The Fe-K line is caught, with the
posterior-predictive AUC reaching $0.96$ at $\sim$10000 counts, but it is harder to
catch at lower counts than on EPIC-pn ($0.76$ against $0.97$ at $\sim$1000 counts,
$0.63$ against $0.76$ at $\sim$100). NICER's effective area at 6.4\,keV is about six
times smaller than at its 1.5\,keV peak ($327$ against $1858$\,cm$^2$,
Fig.~\ref{fig:nicer}\textit{b}), so at matched
total counts fewer photons land near the line and the channel-wise check has less to
work with. The line detectability tracks the effective area at the line energy,
and no per-spectrum check separates the gain shift on either response. The B2
partial-covering family is reported on EPIC-pn only. On NICER the leaked unabsorbed
continuum pushes the flow's posterior well off its training distribution, and the
strongest-leak cell crashed the posterior-predictive replicate simulation with a
Poisson-rate overflow, so only the three weakest faint-level grid points completed
(nine detector cells, in the committed NICER results but not used in any table) and
the family is dropped from the NICER comparison.

\section{Calibration}\label{sec:calibration}

\begin{figure}[t]
\centering
\includegraphics[width=0.92\linewidth]{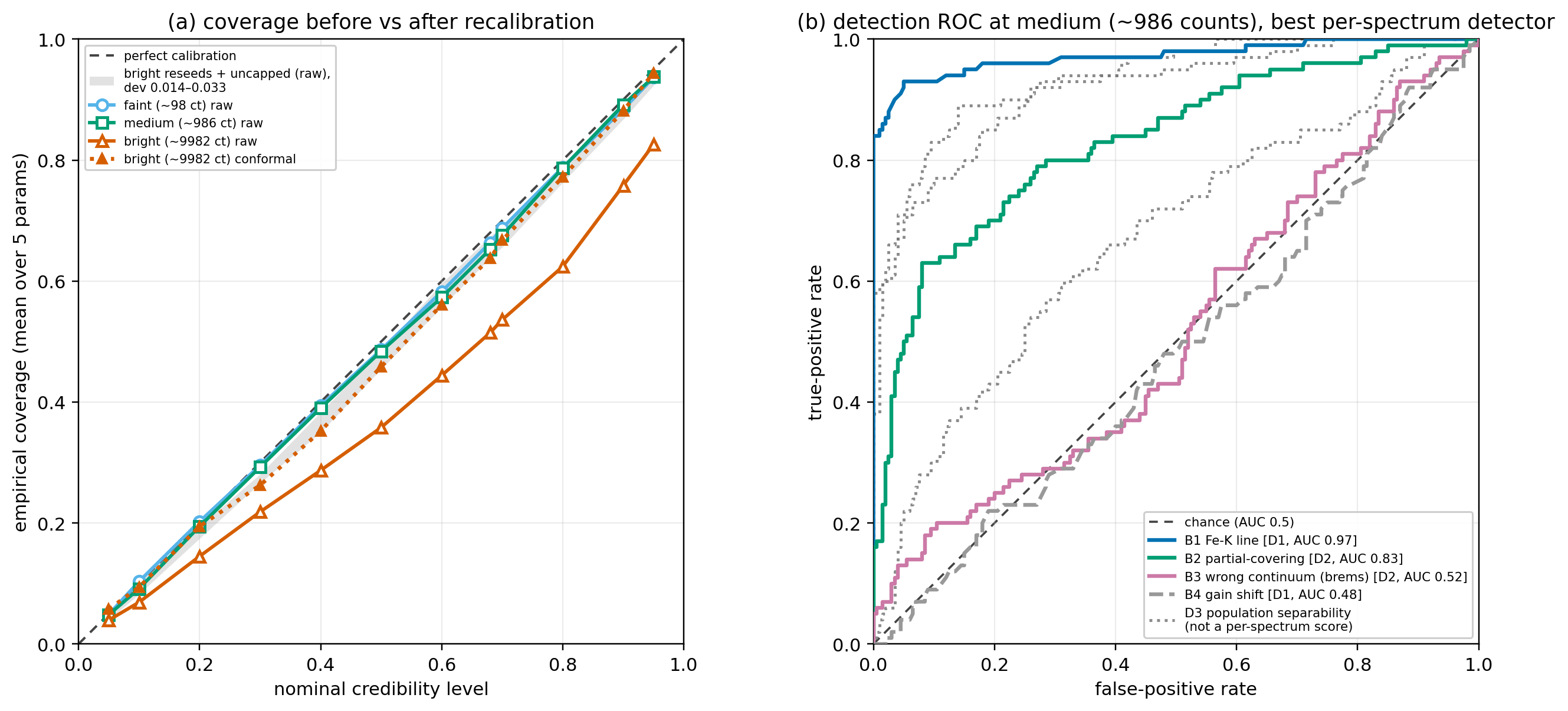}
\caption{\textit{(a)} Empirical coverage, the mean over the five marginals, against
nominal credibility for the miscalibrated production flow. Faint and medium sit on the
diagonal. The bright raw curve sags below it, over-confident at deviation $0.114$, and
split-conformal recalibration (dotted) pulls it back to within $0.031$. The shaded band
is the raw-coverage envelope of the three reseeds and the uncapped retrain of
Table~\ref{tab:gonogo}, each near the diagonal. \textit{(b)} Detection ROC at the medium
($\sim$1000-count) level, the best per-spectrum detector per family at its strongest grid
point, so its AUCs sit at or below Table~\ref{tab:auc}'s entries, which take the best
over the strength grid. B4, the gain shift, sits on the chance diagonal, and D3 (dotted
grey) is overlaid as the population-separability statistic. See
Section~\ref{sec:calibration} for panel \textit{(a)} and Section~\ref{sec:detection} for
panel \textit{(b)}.}
\label{fig:money}
\end{figure}

One trained production flow, the bright one, was miscalibrated, and the calibration suite caught it. It passed every recovery check, with a mean Pearson correlation of $0.84$ between truth and posterior median and all five marginals shrinking monotonically with counts, yet it under-covered badly. Its mean coverage deviation was $0.114$ over the five parameters and twelve nominal levels, against $0.014$ and $0.018$ for the faint and medium flows (Fig.~\ref{fig:money}\textit{a}), and simulation-based calibration (SBC) \citep{talts2018sbc} rejected rank uniformity on all five parameters (KS $p$ below $10^{-13}$ throughout, two of the five underflowing to zero). The flow had hit its 150-epoch training cap with a train/validation gap ($-14.91$ against $-13.36$), so its posteriors narrowed faster than they stayed accurate, a standard over-confidence signature. Split-conformal recalibration \citep{angelopoulos2021} repaired the marginal coverage, $0.114 \to 0.031$, with coverage at nominal $50/68/90$ moving from $0.36/0.51/0.76$ to $0.46/0.64/0.88$; it rescales the one-dimensional marginal intervals and does not address joint coverage. That repair is measured on the bright production flow, and it is the same remap that Section~\ref{sec:fixconformal} runs on the gain-marginalized flow, with the local simulation-based-inference variant of conformal calibration given by \citet{cabezas2025cp4sbi}. Run on a flow whose raw coverage is already near-nominal, that step has nothing to correct (Section~\ref{sec:fixconformal}).

The over-confidence is one production flow's training accident, not the count level. Three reseeds of the bright training, plus one retrain with the epoch cap lifted from 150 to 400, all cover to within $0.014$--$0.033$ (Table~\ref{tab:gonogo}). The uncapped retrain is the cleanest control: same data, cap lifted, and it converged inside the new cap at the lowest deviation in the table, so the $0.114$ traces to a capped, undertrained checkpoint. One reseed hit the same cap and still covered inside that band, so the cap alone does not explain it.

\begin{table}[t]
\centering
\caption{Robustness pass on the bright-level flow. All variants train on the same
$\sim$10000-count data and differ only in the training seed and (for the uncapped
retrain) the epoch cap. Coverage deviation is the mean $|\text{empirical} -
\text{nominal}|$ over five parameters and twelve nominal levels, and SBC $p_{\min}$ is
the smallest per-parameter rank-uniformity KS $p$-value, with the count of failing
parameters in parentheses. Coverage recovers in every variant, and the rank test still
fails in all but one (Section~\ref{sec:calibration}).}
\label{tab:gonogo}
\small
\begin{tabular}{l c c c}
\toprule
Variant & Epochs (cap) & Raw cov.\ dev.\ & SBC $p_{\min}$\\
\midrule
production (orig.) & 151 / 150 (cap hit) & 0.114 & $\approx 0$ (5/5 fail)\\
reseed 101 & 83 / 150 & 0.033 & 0.084 (pass)\\
reseed 202 & 151 / 150 (cap hit) & 0.031 & $1.2\times10^{-8}$ (4/5 fail)\\
reseed 303 & 116 / 150 & 0.022 & 0.016 (2/5 fail)\\
uncapped & 162 / 400 (converged) & 0.014 & 0.028 (1/5 fail)\\
\bottomrule
\end{tabular}
\end{table}

\begin{figure}[t]
\centering
\includegraphics[width=0.98\linewidth]{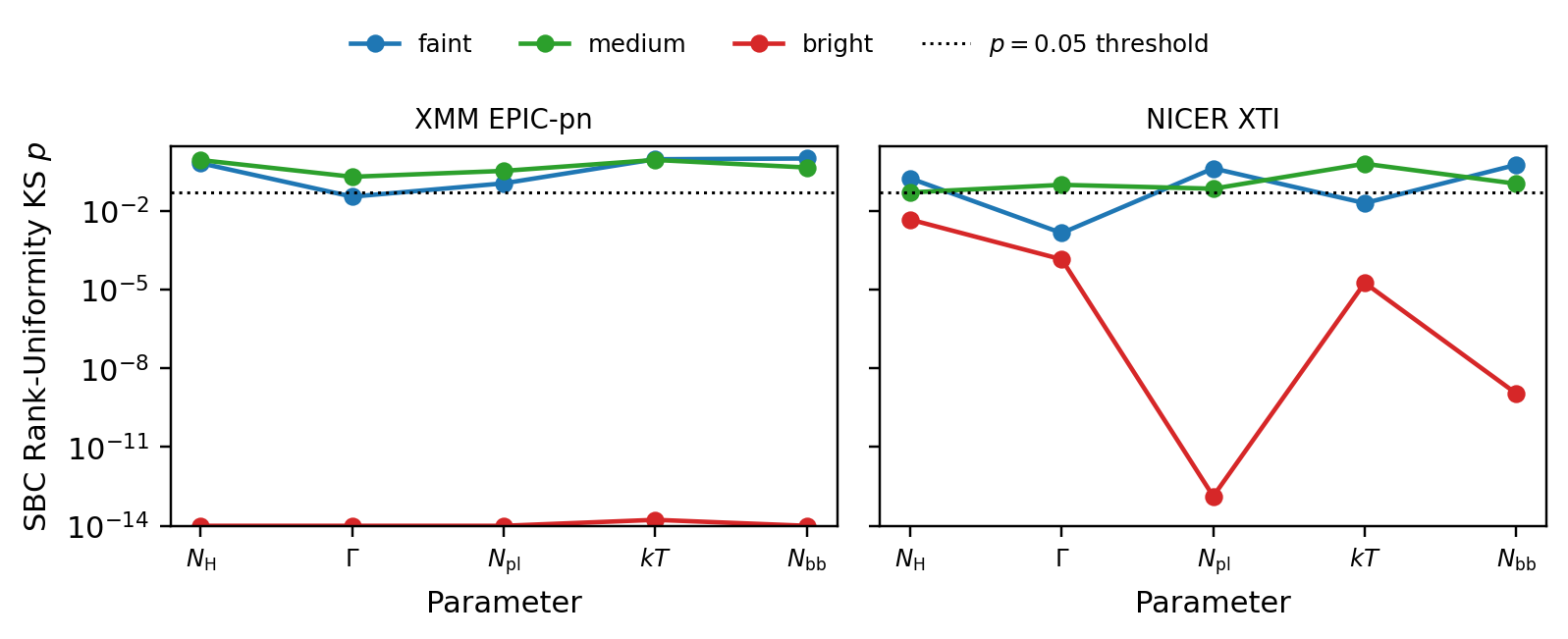}
\caption{Cross-instrument simulation-based calibration. Per-parameter rank-uniformity KS
$p$-value at each count level, for EPIC-pn (\textit{left}) and NICER (\textit{right}),
with the dotted line at $p = 0.05$. The bright ($\sim$10000-count) flow fails on every
parameter on both responses, and faint and medium mostly pass. The EPIC-pn bright flow is
the undertrained production one and fails at the floor. The NICER bright flow is trained
to convergence, at a coverage deviation of $0.011$, and still fails, so the high-count
miscalibration is not one flow's training artifact (Section~\ref{sec:calibration}).}
\label{fig:sbc}
\end{figure}

The rank statistics tell a different story. The power-law normalization fails the SBC rank test at the bright level in four of the five flows, including the uncapped control ($p = 0.028$), while it passes at medium ($0.32$) and faint ($0.11$). The same parameter fails in three of the four other flows, and only one of the three reseeds in Table~\ref{tab:gonogo} passes it. Under the per-flow 5 per cent null that recurrence has probability about $5\times10^{-4}$, rising to $2\times10^{-3}$ after the choice among five parameters, but the five flows share one training set and the uncapped retrain differs from production only in its epoch cap, so treating them as independent draws is generous, and the stronger evidence is cross-instrument. The bright NICER flow, trained to convergence with near-nominal coverage ($0.011$), still fails SBC on all five parameters (Fig.~\ref{fig:sbc}). So the high-count regime carries a residual miscalibration that survives proper training on both instruments, and that the marginal coverage metric is too coarse to see. Conformal fixes the interval widths and leaves the rank structure alone.

\citet{barret2024_i} recover well on a deliberately restricted prior and report passing SBC on that setup. We run the same rank test under a harder wide-prior single-round setup, where it catches a high-count failure the restricted-prior setup does not surface. Recovery metrics do not certify calibration, so a rank test and a coverage test belong on every flow before deployment.

\section{The importance-sampling test}\label{sec:isess}

\citet{barret2026_iii} importance-reweight the NPE posterior against the exact likelihood: draw from the flow, weight each draw by prior times likelihood over flow density, and read the effective sample size of the weights as a quality diagnostic. We report it as an efficiency, the Kish ESS of the weights over a fixed budget of 6000 draws per spectrum. When the flow matches the exact posterior the weights are flat and the ESS approaches the draw count; when it does not, the weights concentrate and the ESS collapses. The scheme puts a per-spectrum guarantee behind every use of the flow. The open question, which \citet{barret2026_iii} defer, is whether the guarantee extends to data the training simulator cannot produce. A systematic that keeps the observed spectrum inside the family of spectra the assumed model can produce leaves the flow and the exact posterior in agreement, just at shifted parameter values, so the weights have nothing to collapse against. So the weights should stay flat under a gain shift, and we test that.

We validated the weights on clean data first. On clean spectra the importance-sampled $\log Z$ agrees with an assumption-free prior Monte Carlo estimate ($10^{6}$ uniform draws, no flow anywhere) to well under a nat at medium counts, and with the benchmark nested-sampling $\log Z$ to under a nat on the spectra both schemes ran exactly, of which there are three at medium, 25 at faint and 15 at bright. Acceptance, the fraction of those 6000 draws that lands inside the prior box, has median 0.98 at medium and 0.92 to 0.94 at bright, with clean and gain-shifted populations within 0.02 of each other, so nothing below is an artifact of differential rejection. The bright acceptance distribution is heavy-tailed, and equally so in both populations. The bright level strains the scheme without breaking it. The bright flow fires its low-ESS alarm (ESS below 0.1 of the draws) on 96.7 per cent of clean spectra. The bright clean ESS fraction (median 0.0095) sits 26 times below the medium value (0.2485), and the clean $\hat{k}$ median at bright is 0.80, past the 0.7 reliability threshold of \citet{vehtari2024psis}. The bright flow is the miscalibrated production flow of Section~\ref{sec:calibration}. We keep that capped flow, because the paired shift measured in Section~\ref{sec:fixbias} does not track a flow's absolute calibration. The medium flow passes the calibration checks and the bright flow fails them, but the paired shift is the same order on both, $+0.02$ in $\Gamma$. Medium is the level the analysis rests on.

Table~\ref{tab:e1auc} gives the ESS-efficiency ROC AUCs, clean against misspecified populations, with bootstrap 95 per cent confidence intervals and permutation $p$-values against 0.5, in two independent seed sets. The line is caught at both levels, AUC 0.683 $[0.615, 0.754]$ at medium and 0.820 $[0.751, 0.882]$ at bright, $p<0.001$ in both seed sets. At medium the gain AUC is 0.458 $[0.388, 0.532]$ ($p=0.257$) and at bright 0.543 $[0.471, 0.615]$ ($p=0.263$), with the second seed set at 0.460 and 0.496. $\hat{k}$, computed on one seed set, behaves the same way on the gain (0.473 medium, 0.540 bright) and sits near chance on the line at medium (0.546); at bright it separates the line (0.809), though the clean $\hat{k}$ median already sits at 0.80 there, so the bright column reads as relative separation only. The line detection at medium comes from the ESS, not from $\hat{k}$. The table starts at medium. The sweep does cover faint, at a median 98 counts, but that arm carries no confidence intervals and no permutation tests, so we quote no AUC from it.

\begin{table}[t]
\centering
\caption{ESS-efficiency ROC AUC (clean vs misspecified), EPIC-pn, bootstrap 95 per cent CIs, permutation $p$ against 0.5, two independent seed sets; each cell compares 150 clean against 100 misspecified spectra. The line is detected at both count levels, and neither seed set separates the gain populations at either level.}
\label{tab:e1auc}
\begin{tabular}{llll}
\hline
level & family & seed set 0 & seed set 1 \\
\hline
medium & B4 (3\% gain) & 0.458 $[0.388,0.532]$, $p=0.257$ & 0.460 $[0.386,0.536]$, $p=0.292$ \\
medium & B1 (Fe-K line) & 0.683 $[0.615,0.754]$, $p<0.001$ & 0.680 $[0.609,0.749]$, $p<0.001$ \\
bright & B4 (3\% gain) & 0.543 $[0.471,0.615]$, $p=0.263$ & 0.496 $[0.420,0.567]$, $p=0.919$ \\
bright & B1 (Fe-K line) & 0.820 $[0.751,0.882]$, $p<0.001$ & 0.795 $[0.732,0.853]$, $p<0.001$ \\
\hline
\end{tabular}
\end{table}

\begin{figure}[t]
\centering
\includegraphics[width=0.6\linewidth]{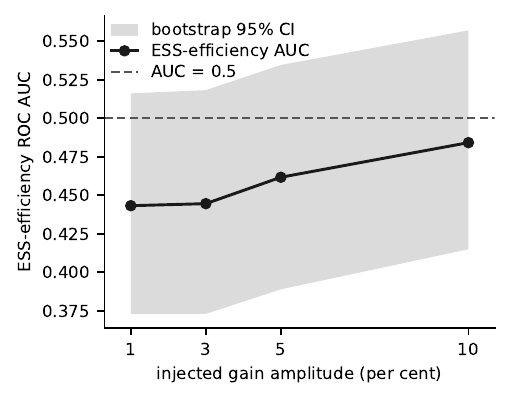}
\caption{ESS-efficiency ROC AUC against injected gain amplitude at medium counts, with bootstrap 95 per cent confidence bands and the chance line at 0.5. Every interval crosses 0.5 from 1 to 10 per cent, and the separation shrinks as the amplitude grows. The zero-amplitude baseline is not plotted here: eight fresh clean populations scored against each other give a mean AUC of 0.4992, so the estimator's null sits at 0.5, and that calibration carries no significance test of its own. The 3 per cent point reads 0.445 against Table~\ref{tab:e1auc}'s 0.458 for the same cell, the same 150 clean spectra scored against a different misspecified draw.}
\label{fig:sweep}
\end{figure}

Blindness at one amplitude could just be lack of power, so we sweep the injected gain at medium counts: 1, 3, 5, and 10 per cent (Fig.~\ref{fig:sweep}). The AUC stays null-consistent, 0.443, 0.445, 0.462, 0.484, every confidence interval crossing 0.5 and every permutation $p \geq 0.13$, and the separation $|{\rm AUC}-0.5|$ falls from 0.057 to 0.016 as the amplitude grows, which is the opposite of what a power-limited detector would do. The zero-amplitude baseline is measured on eight fresh clean populations and given in the caption, and we take it to carry over to these draws. Up to 10 per cent gain, more than three times our stress-test amplitude, the ESS never separates the populations. That also limits the reading that B4 is just a smaller perturbation than B1. B1 sits at one strength while B4 is swept over two decades, and at 10 per cent the 6.4\,keV shift is more than three times the 190\,eV of Section~\ref{sec:setup} and several times the width of the injected line, and it is the least separated point of the four. The weights compare the flow to the exact posterior under the assumed model, and a 3 per cent energy rescale moves the total counts but leaves the spectrum representable, so the free normalizations absorb it and the weights stay flat.

\section{The cross-scheme matrix}\label{sec:matrix}

Table~\ref{tab:matrix} lays the same misspecification families against every scheme, on separate draws of spectra whose sizes the caption gives. The pattern does not depend on which check does the reading. The two flow-side detectors catch the line at medium and bright (D1 at 0.972 and 0.970) and sit at chance on the gain (0.474 to 0.484). The importance-sampling diagnostics do the same (Section~\ref{sec:isess}). The nested-sampling evidence separates the line on the count-controlled comparison, $-67$ nats $[-90, -44]$ at medium and $-892$ $[-1166, -562]$ at bright ($n=6$ each). The gain gets the paired design instead, matched exposure and seed over $n=12$ pairs, and reads $+0.33 \pm 1.37$ nats, consistent with zero ($p=0.812$). All twelve pairs sit at the medium exposure, so the evidence scheme has no gain measurement at bright, and Section~\ref{sec:fixns} is the reason to want one, since bright is where the exact likelihood does carry gain information. We keep the two statistics apart, because the gain shift moves each spectrum's total counts, so any count-referenced evidence statistic confounds the count change with the misfit; the paired design suppresses that confound and the next paragraphs quantify what it was hiding. Faint rows are absent because no misspecified nested-sampling runs exist at faint.

\begin{table}[t]
\centering
\caption{The cross-scheme detectability matrix, 3 per cent gain (B4) and Fe-K line (B1). AUC entries are clean-vs-misspecified ROC AUCs (0.5 = blind), on 200 clean against 100 misspecified spectra for the two flow-side rows (D1, a posterior-predictive check; D2, an embedding-space distance) and 150 against 100 for the importance-sampling rows (IS, importance sampling under the exact likelihood; PSIS, Pareto-smoothed importance sampling); NS entries are evidence differences in nats, count-controlled for B1 ($n=6$) and paired (matched exposure and seed, $n=12$, all at medium exposure) for B4; the paired design was never run on the gain at bright, so that cell carries no entry. No CIs exist for the D1 and D2 AUCs; the $\hat{k}$ row is a single seed set. All rows are EPIC-pn, though the D1 and D2 rows carry no recorded response field and are identified by run configuration.}
\label{tab:matrix}
\begin{tabular}{lcccc}
\hline
scheme & B1 med & B1 bright & B4 med & B4 bright \\
\hline
NPE-PPC D1 (AUC) & 0.972 & 0.970 & 0.482 & 0.474 \\
NPE embedding D2 (AUC) & 0.846 & 0.810 & 0.477 & 0.484 \\
IS ESS-efficiency (AUC) & 0.683 & 0.820 & 0.458 & 0.543 \\
IS PSIS $\hat{k}$ (AUC) & 0.546 & 0.809 & 0.473 & 0.540 \\
NS evidence ($\Delta\log Z$, nats) & $-67\;[-90,-44]$ & $-892\;[-1166,-562]$ & $+0.33 \pm 1.37$ (paired) & \ldots \\
\hline
\end{tabular}
\end{table}

The nested-sampling errors above are measured where they can be. We recompute the per-run evidence error with the thread bootstrap of \citet{higson2018samplingerrorsns} on 13 fresh nested-sampling reruns, and the per-run $\sigma$ has median 0.144 nats, range 0.133 to 0.215. The floor on the paired statistic is 0.060 nats, and that one is an extrapolation. The committed paired spectra cannot be reconstructed from the run record, so we measured the error on five fresh clean-plus-gain pairs at the same medium exposure and transferred it to the mean of all twelve. That holds only as far as those five stand in for the committed ones, and their counts do not reach the committed set's upper end. The empirical SEM of the paired statistic is 1.370 nats, 22.7 times the floor, so the gain null is not nested sampling failing to resolve its own signal, and a margin that wide does not turn on which five pairs stood in. The nested-sampling share of the paired variance is 0.13 to 0.43 per cent, and deleting all of it moves the $p$-value from 0.8116 to 0.8115. The line detections are safe against the per-run error, which on those two line runs is 0.79 per cent of the medium confidence half-width and 0.07 per cent of the bright one, and the floor is smaller still.

That leaves the count-controlled gain residual. Its medium cell is $+13$ nats $[+3, +26]$ at $n=4$, and the positive value reads as an artifact of the regression's high-count anchoring; the reading holds up quantitatively. The sign is wrong for a genuine misfit, since unmodelled structure should lower the evidence, and the size traces to the count mismatch. Across the twelve pairs the shift moves total counts by a median of $+0.6$ per cent, with a per-pair spread from $-10$ to $+3$ per cent. Regressing the paired differences on the count ratio gives $-101.6 \pm 24.2$ nats per $e$-fold of $\log(n_{\rm gain}/n_{\rm clean})$ ($R^2=0.638$, permutation $p=0.0021$), so that per-pair spread carries 64 per cent of the variance of the paired differences, whose standard deviation is 4.75 nats. Adjusting for counts leaves a gain effect of $+0.03 \pm 0.87$ ($p=0.97$). The construction also needs the clean $\log Z$--counts trend ($\log Z = -116.9\,\log_{10}(\mathrm{counts}) + 91.1$, correlation $-0.99$), which only a benchmark that knows the true model can draw. An analyst with a real spectrum cannot build this instrument, and on our benchmark it measures its own count mismatch. We use it nowhere in the gain result.

\citet{cranmer2020frontier} expect misspecification to hit prescribed and implicit likelihoods alike, and \citet{sims2025banter} show a nested-sampling model comparison that scores predictivity in aggregate and so keeps rewarding a model whose components fit wrongly, one component absorbing the residual the other leaves. Ours is the X-ray counterpart, measured on a real detector response and on a systematic the response itself produces.

\section{Gain-marginalized amortized inference}\label{sec:fix}

\subsection{Construction, and the bias it cannot remove}\label{sec:fixbias}

We apply standard nuisance marginalization at training time, adding the gain as a sixth parameter $g \sim \mathcal{U}[0.95, 1.05]$, rescaling the response's energy grid by $g$ in each training simulation, and retraining the flow once per count level (50\,000 simulations, 140 epochs for the medium flow). Nothing changes at inference. The flow conditions on the spectrum and returns a posterior already marginalized over $g$. This is the amortized version of the calibration marginalization long practised in likelihood-based settings \citep{lee2011,xu2014,vitale2020physical}, and the cost is paid once, at training. Inference stays at flow speed, which per-spectrum nested sampling cannot match.

The induced bias is small, so a paired design is needed to see what the fix does and does not do. We evaluate $N=500$ same-parameter pairs, one clean and one gain-shifted realization with common Poisson random numbers, at both count levels. A true 3 per cent shift induces a photon-index bias of $+0.018 \pm 0.006$ under the fixed flow at medium counts, and the gain-marginalized flow returns $+0.020 \pm 0.006$, both errors the paired standard error over the 500 pairs of a single run. Marginalization does not reduce the bias. At bright the seed-averaged bias is $+0.019$, the same order as medium, with $\pm0.005$ the scatter across evaluation seeds, a different statistic from the medium pair. In units of the posterior width this is 0.03 to 0.08$\sigma$ at medium and 0.05 to 0.11$\sigma$ at bright, where $\sigma(\Gamma) \approx 0.20$--$0.22$ because bright $\Gamma$ is degeneracy-limited, not photon-limited. The bias is physical. It flips sign antisymmetrically through an exact zero at $g=1$, grows linearly in $g-1$ and reaches $+0.0178$ at 3 per cent, and a flow-free linearized estimator reproduces its sign and magnitude (medium $+0.0155$, bright $+0.0181$). A pure power law is exactly gain-invariant, so the bias rides the ARF curvature, the band edges, and the blackbody. The one parameter where marginalization appears to help is the bright power-law normalization, whose paired bias halves, from $+0.0063 \pm 0.0011$ to $+0.0031 \pm 0.0010$. The two bright flows differ in training state, so that comparison is confounded.

The gain posterior explains why $\Gamma$ stays biased. At both levels the flow's $g$ marginal returns the prior. Its 90 per cent width is 88 to 90 per cent of the prior's 90 per cent width, which is a different statistic from the standard-deviation ratio Section~\ref{sec:fixns} quotes, and none of the evaluated spectra excludes $g=1$, even under an injected $g=1.03$. With $g$ unidentified, the marginal flow inherits the same degenerate-direction shift the fixed flow had, so the fix cannot repair the point estimate. \citet{vitale2020physical} report the same outcome for gravitational-wave calibration parameters, whose posteriors return their priors. Coverage of $\Gamma$ at the 90 per cent level sits at 85.6 to 88.8 per cent across the four medium cells, the flows' pre-existing calibration level, unchanged by the fix.

The widths stay where they were, and the gain prior predicts almost exactly that. The measured $+0.019$ at 3 per cent gives a local slope $\mathrm{d}\Gamma/\mathrm{d}g \approx 0.6$. A $g$ prior of standard deviation 0.0289 propagated through that slope adds about 0.017 in quadrature to a medium $\sigma(\Gamma)$ of 0.30 to 0.32, a widening of under a per cent. On the same $N=500$ spectra across three independent evaluation seeds, the ratio of gain-marginalized to fixed mean $\sigma(\Gamma)$ at medium runs 0.989 to 0.996, while the median 90 per cent width on the same pairs moves the other way by 0.6 to 1.1 per cent. The sign follows the statistic, and the width is unchanged at the per-cent level. So the calibration uncertainty does not show up in the $\Gamma$ width for a gain prior this wide. Marginalization buys a flow trained through the whole miscalibration family, whose $g$ marginal returns the prior instead of a silent over-confident point. The bright cells move the other way, but their fixed baseline is the epoch-capped flow of Section~\ref{sec:calibration} and the power-law normalization narrows in the same comparison, so nothing there isolates the marginalization. At medium nothing we measured improves under the fix. The argument for it is structural, since no check fires at the amplitudes real detectors hold (Section~\ref{sec:limits}), so an amortized pipeline has to carry the gain in the simulator or not carry it at all, and here carrying it costs nothing we can measure, in posterior width or in inference time.

\subsection{The exact cross-check: where the flow loses gain information}\label{sec:fixns}

\begin{figure}[t]
\centering
\includegraphics[width=0.6\linewidth]{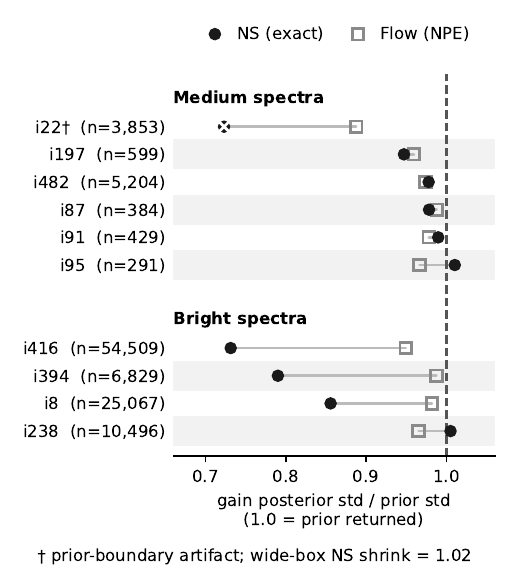}
\caption{Gain-posterior width over prior width for the ten cross-check spectra: exact nested sampling (filled circles) against the gain-marginalized flow (open squares), medium spectra above, bright below. Both sit at or near the prior for the medium spectra, where the gain is unidentified. On the three information-carrying bright spectra nested sampling shrinks to 0.73--0.86 of the prior while the flow barely moves. The medium outlier is i22, a prior-boundary artifact on both sides (crossed nested-sampling point at 0.72, flow point at 0.89); its wide-box rerun returns 1.02 (Section~\ref{sec:fixns}).}
\label{fig:shrink}
\end{figure}

Exact nested sampling with the same sixth dimension re-derives the energy grid from the gain-shifted response on every likelihood evaluation. That construction is the most expensive run in the paper, so it runs on a deliberately small set: ten clean spectra (six medium, four bright), each fit by \textsc{UltraNest} with 400 live points and uncapped $\mathrm{d}\log Z < 0.5$, against the gain-marginalized flow on the same spectra. Flow draws use in-prior rejection with clipping, the convention behind every capture number below. Shrink is the posterior standard deviation over the prior standard deviation, and capture is $(1-\mathrm{shrink_{flow}})/(1-\mathrm{shrink_{NS}})$, the share of the exact run's gain constraint the flow reproduces. Shrinks are quoted against the continuous prior standard deviation 0.0289, while Fig.~\ref{fig:shrink} scales the flow points by the 200-atom $g$ grid the flow samples, 0.5 per cent wider, so captures recomputed from the figure run 1.1 to 1.7 times the quoted values. The comparison measures disagreement, not correctness; against the injected truth the two are statistically indistinguishable (nested-sampling RMS $|z| = 0.880$ against the flow's 1.038, Wilcoxon $p=0.20$).

At medium counts the exact posterior agrees with the flow that $g$ is unidentified. Nested-sampling $g$-shrinks sit at 0.95 to 1.01 of the prior width for five of the six medium spectra, and a Fisher estimate for the medium population predicts a shrink of 0.987, so the physics requires the flow's prior-return at medium. The sixth spectrum, i22, shrank to 0.723 and looked like a medium-level counterexample, but it is a prior-boundary artifact. Its $kT \propto g$ blackbody degeneracy saturates at the $kT$ prior ceiling exactly where nested sampling cuts, the exact likelihood is flat to 0.85 nats across the full $g$ prior, and a rerun with the physics box widened and the $g$ prior unchanged returns the shrink to 1.022. The rule was pre-registered before that rerun: a value near 1.0 confirms the artifact, near 0.72 reopens it.

Bright behaves differently. Three of the four bright spectra carry gain information. The finite-shift Fisher error is $\sigma_{\rm eff} = \Delta g/\sqrt{R}$, with $R$ the residual power a gain shift $\Delta g$ leaves after the other five parameters absorb what they can. For the three it reads 0.037, 0.050, and 0.039 against a prior standard deviation of 0.0289. On those three the exact posterior shrinks $g$ to 0.731, 0.790, and 0.856 of the prior, while the flow captures 0.163, 0.04, and 0.07 to 0.11 of that shrink (Fig.~\ref{fig:shrink}). The 0.163 is a 16-seed mean, so it does not reproduce exactly from the single shrink values quoted here. It belongs to i416, which carries $54\,509$ counts, five times the nominal bright median and in the sparse upper tail of the flow's own training counts. We use $\sigma_{\rm eff}$ ordinally, only to screen which spectra carry gain information at all. It is not a width prediction. It does not track the individual shrinks case by case, and the three values above are not even monotone in them. The soft band explains it. Every continuum component on its own is exactly gain-degenerate; \texttt{tbabs} and the band edges break the degeneracy, so the gain information lives in the absorbed soft-band counts below 1.2\,keV (Spearman $+0.964$ with $N_{\rm H}\sqrt{n_{\rm soft}}$ across the set), which is why the bright spectrum with 6 per cent soft counts shows no shrink. We checked that the shrinks are robust to the sampler. A different sampler family reproduces the i394 and i416 shrinks to 2.4 and 0.9 per cent, and doubling its steps moves them by at most 1.1 per cent. A classifier two-sample test between the exact and flow posteriors in the full six dimensions reads 0.91 for i394 and 0.92 for i416, and doubling the sampler's steps moves it by under 0.01. Most of that separation sits in the continuum parameters, though; the gain dimension on its own reads 0.60 to 0.64.

So the scope of ``unidentifiable'' follows the absorbed soft-band counts, and count level is only the population-level correlate. Medium draws rarely carry enough soft-band signal, and at bright a Fisher screen over the prior population puts gain information in roughly half the draws. At medium the prior return is physics, and the flow is faithful. On the three bright spectra here that do carry gain information the flow delivers at most a sixth of what the exact run extracts, so the amortization loses most of the gain constraint the data supply. i394 shows it most clearly. Its $N_{\rm H}$, $\Gamma$, and normalization offsets lie along one degeneracy direction (effective dimension 1.11 of 3), the direction is corroborated by the injected truth, and the mean 68 per cent-interval intersection-over-union against the exact posterior is 0.28.

\subsection{Conformal recalibration}\label{sec:fixconformal}

The same split conformal recalibration \citep{angelopoulos2021} rescales interval widths and by construction cannot move the point estimate, so the $\Gamma$ bias passes through untouched, and at this shift strength there is nothing left to repair either. Raw coverage is already near-nominal, but on held-out sets the conformal step makes the coverage deviation slightly worse (mean absolute deviation $0.0147 \to 0.0199$ on gain-shifted spectra, $0.0136 \to 0.0196$ on a disjoint clean control; the second replication repeats both within 0.001). That reading is specific to this medium gain-marginalized flow, whose raw deviation is already 0.0136 to 0.0147. Split conformal does repair the marginal coverage of the miscalibrated production flow of Section~\ref{sec:calibration}, which starts at 0.114. The same experiment replicates the bias with an independent design: the shift-attributable $\Gamma$ bias is $+0.021 \pm 0.008$ precision-weighted (stratified permutation $p=0.010$) and $+0.024 \pm 0.015$ unweighted ($p=0.114$), about $+0.07$ to $0.08\sigma$, consistent with the paired measurement and underpowered on its own.

\section{Realistic amplitude and limits}\label{sec:limits}

The stress-test framing cuts the other way too, so we also measure at the amplitudes real detectors hold (0.1 to 0.3 per cent, Section~\ref{sec:intro}). The ESS detector at medium counts, swept at 0.1, 0.2, 0.3, and 0.5 per cent injected gain, reads AUC 0.457, 0.437, 0.505, and 0.455, every confidence interval crossing 0.5 and the worst-case $p$ at 0.094. Pooling four independent seed sets per amplitude gives 0.495, 0.481, 0.505, and 0.480 ($p = 0.80$, 0.32, 0.79, 0.27), with no trend. One seed set alone showed 0.53 to 0.59, with two points at nominal $p<0.05$. Against the size-matched null standard error of 0.0373 those two sit at $z=2.16$ and 2.42, and they are two of the 16 swept cells, with a third nominal hit in another seed set pointing below 0.5. Neither replication seed set reproduces the elevation, one of them sitting below 0.5 at all four amplitudes, so with the pooled means above we read it as draw noise. The null at realistic amplitude is measured directly, and the null-consistent 1 to 10 per cent sweep already implied it.

Detection signal grows with amplitude, so the same scaling downgrades every claim that a check fired. The counts a detection needs scale as the inverse square of the amplitude, so a check that could barely fire at 3 per cent near $10^{3}$ to $10^{4}$ counts would need of order $10^{5}$ to $10^{7}$ counts at the 0.1 to 0.3 per cent level, beyond essentially any real spectrum in this class. We make no detection claim at realistic amplitude for any scheme, and the fix of Section~\ref{sec:fix} reads as marginalize-only there. The retrained flow will not widen anything at 0.1 to 0.3 per cent either, since the widening predicted for the much wider ${\pm}5$ per cent training prior is already under a per cent (Section~\ref{sec:fixbias}).

Marginalization also needs the analyst to posit the nuisance. The retrained flow corrects only the systematic it models. Replacing the benchmark diagnostics with that flow narrows the ground-truth dependence without removing it; \citet{sims2025banter} state the same limitation for their validation null test, which covers only the nuisance structure its validation data contain, and it applies here unchanged. The sub-percent sweep is scoped to the gain family at medium counts on EPIC-pn. Everything here also rests on the one five-parameter source model of Section~\ref{sec:setup}. The free normalizations and the continuum shape reabsorb the rescale, so a fit with less continuum freedom, or one restricted to a narrow band, would leave a residual this benchmark does not have, and the blindness need not carry over to it. The benchmark is entirely simulation-based, and no observed spectrum is fit. The forward model carries no background component, so a real spectrum adds a background term the model would have to absorb or fit. B4 is a slope-only rescaling, and a real gain error can also carry an offset, which we do not inject.

\section{Why the blindness happens}\label{sec:mechanism}

The pattern has a geometric reading. A misspecification is invisible to a check built on the fitted model when the data perturbation it produces lies inside the span of the model's parameter Jacobian: the fit reabsorbs it, the parameters shift along the degenerate direction, and no residual survives for any diagnostic to score. This is Fisher-information geometry of the sloppy-model kind \citep{transtrum2015sloppy}, and the complement of the nuisance-hardened score compression of \citet{alsingwandelt2019}. The gain shift is the in-span case almost everywhere, and the small out-of-span remainder lives in the absorbed soft band (Section~\ref{sec:fixns}), which is also where the exact posterior finds its gain information at bright counts. The 6.4\,keV line is the out-of-span case, a localized feature no continuum parameter can build, so every scheme sees it. The evidence columns need care beyond this local picture, since a prior-integrated, Occam-weighted quantity is not governed by a local Jacobian; we observe the blindness there (Section~\ref{sec:matrix}), but we do not derive it. The photon-index bias is the same reabsorption seen from the parameter side: the fit moves along the degenerate direction by an amount set by the ARF curvature and the band edges, $+0.02$ in $\Gamma$ at 3 per cent, and marginalizing an unidentified $g$ cannot undo a shift the data cannot see.

\section{Conclusions}\label{sec:conclusions}

\begin{itemize}
\item Every check we ran missed the 3 per cent gain shift, a factor 2.4 to 14.4 above the EPIC-pn energy-scale accuracy. The flow-side AUCs run 0.474 to 0.484, the ESS AUCs 0.44 to 0.54, and the paired evidence is $+0.33 \pm 1.37$ nats at medium exposure, with a sampling floor 22.7 times below that error bar, extrapolated from five proxy pairs. The 6.4\,keV line is the control that says the schemes work at all, and it lands at 0.97, 0.68 to 0.82, and $-67$ to $-892$ nats on the count-controlled comparison. Sweeping the injected gain from 1 to 10 per cent never separates the populations, and neither does the sub-percent sweep at 0.1 to 0.3 per cent, which we ran for the effective sample size at medium counts.
\item The shift is not harmless. The photon-index bias runs 0.03 to 0.11$\sigma$, $+0.02$ in $\Gamma$ at both $10^{3}$ and $10^{4}$ counts, sign and size matched by a flow-free linearized estimator. Gain marginalization does not remove the bias, because the data leave $g$ unidentified, and a ${\pm}5$ per cent gain prior does not measurably widen the posterior either. The fix buys the modelled nuisance and a $g$ marginal that returns the prior, at amortized cost. The bright power-law normalization bias also halves, a modest ${\sim}2\sigma$ difference, but its fixed baseline is the undertrained bright flow of Section~\ref{sec:calibration}, so that comparison does not isolate the marginalization.
\item Ten spectra fit with exact gain-marginalized nested sampling scope the fix: the flow is faithful where the data carry no gain information (predicted shrink 0.987 for the medium population) and lossy where the absorbed soft band does carry it (exact shrinks 0.73--0.86 of prior against a flow capture of at most 0.163 on the continuous-prior convention).
\item At catalog scale the rule for X-ray SBI is to marginalize over any energy-scale nuisance the calibration cannot exclude. One retrain per count level buys it, and the nuisance becomes part of the model.
\end{itemize}

{\small
\section*{Data availability}

The code, configurations, and committed outputs behind the numbers in this paper, including the detection-benchmark ROC grids, the flow calibration summaries on both responses, the conformal summaries on EPIC-pn, the importance-sampling sweeps and their null calibration, the paired and count-controlled nested-sampling summaries with the thread-bootstrap error-bar outputs and their code, the gain-marginalized cross-check runs, and both figure scripts, are in the \texttt{sbi-xray-calibration} repository\footnote{\url{https://github.com/WizardEternal/sbi-xray-calibration}}. Data generation and evaluation reproduce from configuration and fixed seeds; training is seeded but not bit-deterministic.

\bibliographystyle{plainnat}
\bibliography{successor}
}

\end{document}